\documentclass[11pt]{article}
\usepackage{epsfig}
\def\bc{\begin{center}}
\def\ec{\end{center}}
\newcommand{\be}{\begin{equation}} 
\newcommand{\ee}{\end{equation}} 
\newcommand{\bea}{\begin{eqnarray}} 
\newcommand{\eea}{\end{eqnarray}} 
\newcommand{\agg}{A_{\gamma\gamma}}

\newcommand{\azz}{A_{\scriptscriptstyle ZZ}}
\newcommand{\agz}{A_{\scriptscriptstyle \gamma Z}}
\newcommand{\azg}{A_{\scriptscriptstyle Z \gamma}}
\newcommand{\pigg}{\Pi_{\gamma\gamma}}
\newcommand{\piggp}[1]{\mbox{$\Pi^{\mbox{\scriptsize{(#1)}}}_{\gamma\gamma}$}}
\newcommand{\piggpms}[1]{
         \mbox{$\hat{\Pi}^{\mbox{\scriptsize{(#1)}}}_{\gamma\gamma}$}}
\newcommand{\mt}{m_{t}^2}
\newcommand{\mw}{m_{\scriptscriptstyle W}^2}
\newcommand{\mzs}{m_{\scriptscriptstyle Z}^2}
\newcommand{\mz}{m_{\scriptscriptstyle Z}}
\newcommand{\mh}{m_h^2}

\newcommand{\ehs}{\mbox{$\hat{e}^{2}$}}
\newcommand{\alphah}{\mbox{$\hat{\alpha}$}}
\newcommand{\ms}{\mbox{$\overline{ MS}$}}

\newcommand{\Ord}[1]{\mbox{${\cal O}\!\left(#1\right)$}}
\newcommand{\Mvariable}[1]{#1}

\newcommand{\hw}{h_{\scriptscriptstyle W}}
\newcommand{\tw}{t_{\scriptscriptstyle W}}
\newcommand{\zt}{z_{t}}
\newcommand{\htt}{h_{t}}
\newcommand{\Dah}{\Delta \alpha^{(5)}_{\rm had}(\mzs)}

\catcode`\_=11 
\def\c_ncel#1#2{\ooalign{\lower.2ex\hbox{$\hfil#1\mkern.7mu/\hfil$}\crcr$#1#2$}} 
 
\catcode`\_=8 
\catcode`\_=11 
\def\c_ncel#1#2{\ooalign{\lower.2ex\hbox{$\hfil#1\mkern.7mu/\hfil$}\crcr$#1#2$}} 
 
\catcode`\_=8 
\catcode`\_=11 
\def\c_ncel#1#2{\ooalign{\lower.2ex\hbox{$\hfil#1\mkern.7mu/\hfil$}\crcr$#1#2$}} 
 
\catcode`\_=8 
\catcode`\_=11 
\def\c_ncel#1#2{\ooalign{\lower.2ex\hbox{$\hfil#1\mkern.7mu/\hfil$}\crcr$#1#2$}} 
 
\catcode`\_=8 
\catcode`\_=11 
\def\c_ncel#1#2{\ooalign{\lower.2ex\hbox{$\hfil#1\mkern.7mu/\hfil$}\crcr$#1#2$}} 
 
\catcode`\_=8

\begin{document}

\thispagestyle{empty}
\setcounter{page}{0}
\bc
\hfill{RM3-TH/03-07}\\
\hfill{IFUM-764/FT}
\ec

\vspace{1.7cm}

\bc
{\bf \LARGE Two-loop renormalization of the electric charge\\
 in the Standard Model}
\ec

\vspace{1.4cm}

\bc
{\Large \sc Giuseppe Degrassi~$^{a}$ and Alessandro Vicini~$^{b}$} \\
\vspace{1.2cm}

${}^a$
{\em Dipartimento di Fisica, Universit\`a di Roma Tre,\\
INFN, Sezione di Roma III, Via della Vasca Navale 84, I--00146 Rome, Italy}
\vspace{.3cm}

${}^b$
{\em  Dipartimento di Fisica ``G. Occhialini'', 
Universit\`a degli Studi di Milano,\\
INFN, Sezione di Milano,
Via Celoria 16, I--20133 Milano, Italy}\\

\ec

\vspace{0.8cm}

\centerline{\bf Abstract}
\vspace{2 mm}
\begin{quote} \small
We discuss the renormalization of the electric charge  at the two-loop level
in the Standard Model of the electroweak interactions.
We explicitly calculate the expression of the complete on-shell two-loop 
counterterm
using the Background Field Method
and discuss the advantages of this computational approach.
We consider the related quantity $\ehs(\mu)$,
defined in the $\ms$ renormalization scheme and present numerical
results for different values of the scale $\mu$.
We find that the full 2-loop electroweak corrections contribute more
than 10 parts in units $10^{-5}$ to the $\Delta \alphah (\mzs)$
parameter, obtaining $\alphah^{-1}(\mz)= 128.12 \pm 0.05$ 
for $\Dah =0.027572 \pm 0.000359$. 
\end{quote}
\vfill
\newpage
\section{Introduction}
The very high experimental precision reached at LEP and prospected at
TESLA with the GigaZ option, requires a corresponding theoretical
effort to provide accurate predictions. Inclusion of higher
order effects and a very precise knowledge of the input parameters
of the electroweak Standard Model (SM) are necessary ingredients of precision 
physics. Among the three basic input
parameters usually employed, namely $\alpha,\, G_\mu$ and $\mz$,
the fine structure constant defined at zero momentum transfer,
$\alpha (0)$, is the most precise one with a relative error
of 3.7 parts per billion. However, for physics at high momentum transfer,
like physics at the $Z$-resonance, the use of an effective coupling 
defined at the relevant scale is more appropriate, e.g.\ for the $Z$-resonance 
$\alpha (\mz)$  is more adequate than $\alpha (0)$. 

In pure QED the
natural definition of an effective QED coupling at the scale $\sqrt{s}$
\bea
\label{alphaqed}
\alpha (s) &=& \frac{\alpha}{1 - \Delta \alpha (s)} \\
\label{deltaal}
\Delta \alpha (s) &=&4 \pi \alpha \, {\rm Re}\,\left[ \pigg (s) - 
                          \pigg(0) \right]~,
\eea
is given in terms of the photon vacuum polarization function evaluated
at different scales.

In the full SM, the bosonic contribution to the photon vacuum
polarization at high momentum transfer is, in general, not gauge-invariant.
Thus it cannot be included in a sensible way in Eq.(\ref{alphaqed}).
Eq.(\ref{alphaqed}) with only the fermionic contribution included
is a good effective coupling at the $\mz$ scale.  However, for
energy scales much higher than $\mz$, that will be tested by the future 
accelerators, an effective 
QED coupling that  takes into account  also the bosonic contributions
can be considered.

A different definition of a QED effective coupling can be obtained
by considering the \ms\ QED coupling constant at the scale
$\mu$ defined by
\be
\alphah (\mu) = \frac{\alpha}{1 + (2 \hat{\delta e}/e)}~.
\label{alfamu}
\ee
Eq.(\ref{alfamu}) is expressed in terms of the finite part of the on-shell electric 
charge counterterm (i.e. with the dimensional regularization pole subtracted), 
which is  gauge-invariant quantity that includes
both fermionic and bosonic contributions. In the Background Field Method 
(BFM), as it will be discussed in detail in section 3,
the counterterm is given only by the photon vacuum polarization diagrams,
evaluated at $q^2=0$.
At the one-loop level the electric charge renormalization has been discussed 
in \cite{sir80,marsir}. 

In this paper we present explicit results for  the electric charge counterterm
including all second order ${\cal O}(\alpha^2)$ electroweak
corrections. Our calculation is performed employing the BFM framework.  
The issue of the two-loop renormalization of the electric charge in the SM
was already addressed in the usual  $R_{\xi}$ gauge quantization scheme
by several papers discussing the two-loop contributions to the
$m_{\scriptscriptstyle W}$-$\mz$ interdependence \cite{Dr2l}.
Our  calculation provides the necessary ingredients to define and 
evaluate numerically the effective parameter $\ehs(\mu)$, 
which is a fundamental quantity in all precision tests of the SM.

The paper is structured in the following way. 
In Section \ref{structure} we outline the calculation of the Thomson
scattering amplitude, which allows to define the electric charge
counterterm, and present the 1-loop result in the SM.
In Section \ref{secBFM} we discuss the main differences between the
usual  $R_{\xi}$ gauge quantization scheme   and the
approach offered by the BFM, that makes
manifest the possibility of a Dyson summation also for the bosonic
contribution. 
In Section \ref{results} we present the results of our calculation of
the Thomson scattering amplitude at the 2-loop level and comment on
the checks that we made.
In Section \ref{secehs} we discuss in detail the $\ms$ parameter
$\ehs(\mu)$, 
present numerical results for this parameter for different values
of the scale $\mu$, and discuss the relevance at $\mu=\mz$
of the contributions we have computed. Finally, we comment on the variation
of the 95 \% upper limit on the Higgs boson mass induced by our new
result on $\alphah (\mz)$.

\section{Structure of the calculation}
\label{structure}
\begin{figure}[h]
\centering
\epsfig{file=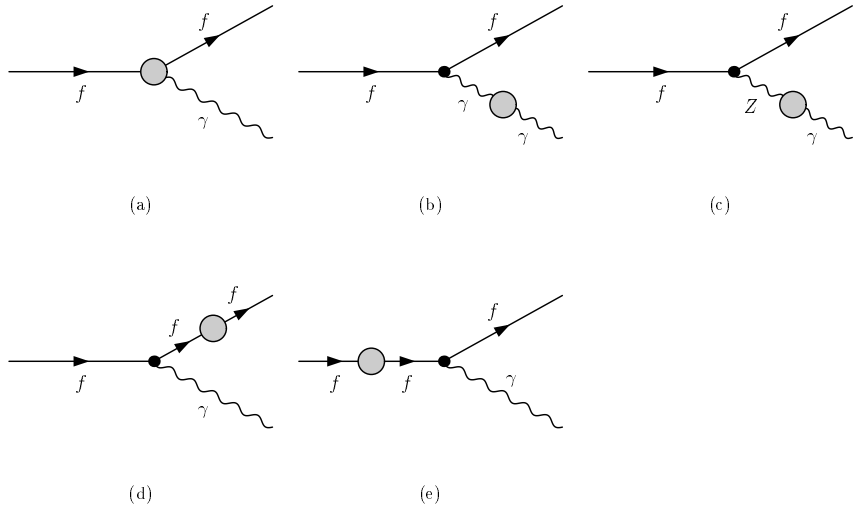,height=8cm,width=12cm,
        bbllx=150pt,bblly=300pt,bburx=600pt,bbury=600pt}
\caption{The 
diagrams of the Thomson scattering
 }
\label{fig1}
\end{figure}

The electric charge is defined in terms of  Thomson   scattering,
namely of the scattering of a fermion off a photon of vanishingly
small energy. The diagrams that describe this process in the SM 
can be depicted symbolically as in fig. \ref{fig1}. 

As it is well known in pure QED the $\gamma$--$Z$ mixing 
diagram (\ref{fig1}c) is 
absent while the Ward--Takahashi identity ensures the cancellation of the 
vertex  contribution (\ref{fig1}a) against the wave function renormalization 
of the fermion (\ref{fig1}d, \ref{fig1}e) such that the relation between the 
bare charge $e_0$ and the conventional renormalized charge~$e$ can be written,
via Dyson summation, as:
\be
    e^2 = {e^2_0 \over 1 -e^2_0\, \piggp{f}(0) }
\label{eq1}
\ee
where $\piggp{f}(0)$ is the  fermionic QED vacuum-polarization function
evaluated at $q^2 = 0$.

We write in general a vector boson (V) self-energy
as
\be
\label{se}
\Pi^{\mu\nu}_{VV}(q^2)= A_{VV}(q^2) g^{\mu\nu} + B_{VV}(q^2) q^{\mu} q^{\nu},
\ee
employing the  convention that the photon vacuum polarization function
is related to the transverse part of its self-energy by
\be
 \agg (q^2) = q^2 \,e^2_0\,\pigg (q^2)~.
\label{eq2}
\ee

The discussion of the  Thomson scattering in the full SM when the theory is 
quantized employing the  conventional linear $R_{\xi}$ gauge-fixing 
procedure \cite{FLS}
differs from the QED case. We recall that in the $R_{\xi}$ gauges the 
classical Lagrangian is supplemented by a gauge fixing function of the form 
\bea
\label{RFG}
{\cal L}^{g.f.}_{R_{\xi}} &=&
-\frac{1}{2 \xi}
\left(
F_{\gamma}^2+F_Z^2 + 2 F_+ F_-
\right)\\
F_{\pm}&=&
\partial^{\mu} W^{\pm}_{\mu} \mp i\,\xi\, m_{\scriptscriptstyle W} \phi^{\pm}
\nonumber\\
F_Z&=&
\partial^{\mu} Z_{\mu} -  \xi\,\mz \chi \nonumber\\
F_{\gamma}&=&
\partial^{\mu} A_{\mu}\nonumber
\eea
that cancels, at the tree-level, the mixing between the vector and scalar 
fields. 
In Eq.(\ref{RFG}) $\phi$ and $\chi$ are the unphysical counterparts associated
to the $W$ and $Z$ bosons.
 
In the SM the radiating fermion couples both to the photon and
the $Z$ currents ($J_{\gamma}^\mu,J_Z^\mu$), the latter via the 
$\gamma$--$Z$ mixing diagram, fig.(\ref{fig1}c). Furthermore, 
the theory does not satisfy a QED-like Ward identity, namely
the sum of diagrams 
(\ref{fig1}a, \ref{fig1}d, \ref{fig1}e) does not add anymore to zero. Instead
they  come out proportional to the third current of the weak isospin
$J_3^\mu \equiv 2( J_Z^\mu + s^2 J_\gamma^\mu)$ with  $s^2 \equiv 
\sin^2 \theta_W$,
so that the part of $J_3^\mu$ proportional to the $Z$ current cancels the
contribution coming from the $\gamma$--$Z$ mixing in order to obtain a 
result only proportional to the photonic current. The final result is 
constituted by the total photon self-energy contribution (fermionic plus 
bosonic\footnote{We classify  as fermionic any self-energy diagram that 
contains at least one fermionic line while all the others are indicated as
bosonic.}) plus the  vertex part from diagrams (\ref{fig1}a, \ref{fig1}d, 
\ref{fig1}e) proportional to $J_\gamma^\mu$. At the one loop level
we have \cite{sir80}
\be
    e^2 =
    e^2_0 \left\{
                1 + e^2_0\, \piggp{f}(0) - \frac{7e^2_0}{8\pi^2}
        \left[
         \frac{1}{n-4} + \ln\frac{m_{\scriptscriptstyle W}}{\mu} - \frac{1}{21}
        \right]
                \right\} \\
    \equiv 
    e_0^2 \left( 1 + \frac{2 \delta e^{(1)}}{e}  \right),
\label{Oneloop}
\ee 
where $\delta e^{(1)}$ is the on-shell one-loop electric charge counterterm.
In Eq.(\ref{Oneloop}) the last term in the curly bracket represents 
\Ord{e^2_0} bosonic
contributions to the charge renormalization and in the $\xi=1$ Feynman
gauge  3 out of the 7 parts come
from the bosonic contribution to the photon self-energy while
the remaining 4 are from the vertex diagrams. 
In Eq.(\ref{Oneloop})
$n$ is the dimension of the space-time and and $\mu$ is a rescaled
`t Hooft mass according to
\be
        \mu \rightarrow {\mu\,e^{\gamma/2}\over (4\pi)^{1/2}}.
\label{E5b}
\ee
The factor $e^{\gamma/2}(4\pi)^{-1/2}$ is appended to the usual
`t Hooft mass  in order to cancel some numerical constants that are 
an artifact of dimensional regularization~\cite{Rmu}.
We notice that, because of the presence of non-vanishing vertex 
contribution, the possibility of a Dyson summation like Eq.(\ref{eq1})
in the SM with linear gauge fixing is not manifestly evident. 

In general  the renormalization of the electric charge in the
SM with linear gauge-fixing requires  the evaluation of the full set of 
diagrams of fig. \ref{fig1} and beyond one-loop level it can become quite
complicated although the analysis could be somewhat simplified with
an appropriate use of the relevant Ward identity (see section 4). However,
the problem cannot be reduced to the calculation of 
just the photon vacuum polarization as in pure QED because of 
the lack of a QED-like Ward identity. 

\section{Background-field method analysis}
\label{secBFM}
As it is well known, in a gauge theory  the choice of  a gauge in order 
to quantize the theory can spoil in the intermediate steps the 
original gauge symmetry of the lagrangian that is actually restored at the 
end when physical processes are considered. This is what actually happens 
when the SM is  quantized with the linear gauge-fixing function of 
Eq.(\ref{RFG}).
The BFM \cite{BFM,abbott} is a technique for 
quantizing  gauge  theories that avoids the complete
explicit breaking of the gauge symmetry.  
One of the salient features of this approach is that all fields are
splitted in two components: a classical background field $\hat{V}$
and a quantum field $V$ that appears only in the loops.
The gauge-fixing procedure is achieved through a non linear term in the
fields that breaks the gauge invariance only of the quantum part of the 
lagrangian,
preserving the gauge symmetry of the effective action with respect to the 
background fields. As a result the background field Green functions
satisfy simple QED-like Ward identities. 

The application of the BFM to the SM was discussed in Ref.\cite{ddw}.
A suitable generalization  of the gauge-fixing term of Eq.(\ref{RFG}) to the
BFM that retains the gauge invariance of the action under background
field transformation can be written as \cite{BFGGF}:
\bea
\label{BFG}
{\cal L}^{g.f.}_{BFM} &=&
-\frac{1}{2 \xi_Q}
\left(
G_{\gamma}^2+G_Z^2 + 2 G_+ G_-
\right)\\
G_{\pm}&=&
\partial^{\mu} W_{\mu}^{\pm}\mp i\, \xi_Q\,m_{\scriptscriptstyle W} \phi^{\pm} 
\pm i  
\left(e {\hat A}_{\mu} -g \,c {\hat Z}_{\mu}\right) W_{\mu}^{\pm}  
\pm i  
\left(e A_{\mu} - g \,c Z_{\mu}\right) {\hat W}_{\mu}^{\pm}
\mp\nonumber\\ 
&&~\mp \frac{i}2 g \xi_Q 
\left[
( {\hat H} \mp i {\hat\chi}) \phi^{\pm}
-(H\mp i\chi) {\hat\phi}^{\pm}
\right]
\nonumber\\
G_{Z}&=&
\partial^{\mu} Z_{\mu}- \xi_Q\, \mz \chi 
+ i \,g\,c 
\left({\hat W}_{\mu}^+ W^{-\mu} - W_{\mu}^+ {\hat W}^{-\mu} \right)
+i g \xi_Q 
\frac{c^2-s^2}{2 c\, s}
\left({\hat \phi}^+ \phi^{-} - \phi^+ {\hat \phi}^{-} \right)~+\nonumber\\
&&~+~~g \xi_Q 
\frac{1}{2 c}
\left({\hat \chi} H -  {\hat H} \chi - v \chi \right)
\nonumber \\
G_{\gamma}&=&
\partial^{\mu} A_{\mu}\nonumber
+i e \left({\hat W}_{\mu}^+ W^{-\mu} - W_{\mu}^+ {\hat W}^{-\mu} \right)
+i e \xi_Q\left({\hat \phi}^+ \phi^{-} - \phi^+ {\hat \phi}^{-} \right)
\eea
where $g$ is the $SU(2)$ coupling, $c \equiv \cos \theta_W$, $\xi_Q$ the
quantum gauge parameter and $H$ the physical Higgs field.

The invariance of the effective action under the relevant background gauge
transformation of the background fields allows to write  identities that
have a simpler structure of  the conventional Slavnov-Taylor
identities and in general  do not involve ghost fields. In particular,
for the two and three point functions involving the photon 
the following identities hold to all orders in perturbation theory.
\bea
\label{QEDWard}
q^{\mu} \Gamma^{\gamma {\bar f} f}_{\mu}(q,{\bar p},p) &=& 
-e Q_f
\left[
\Sigma_f({\bar p}) - \Sigma_f(-p)
\right]
\\
\label{GGLzero}
B_{\gamma\gamma}(0) &=& 0\\
\label{GZLzero}
B_{\gamma Z}(0) &=& 0
\eea
where  $\Gamma^{\gamma {\bar f} f}_{\mu}$
is the three-point function photon-fermion-antifermion, 
$\Sigma_f$ is the fermion two-point function, $q= {\bar p} + p$
the photon momentum 
and $Q_f$ is the charge of the fermion $f$ in units $e$. 
Eq.(\ref{QEDWard}) is the usual QED Ward identity.
Eqs.(\ref{QEDWard}) and (\ref{GZLzero})
are not true in the conventional  $R_\xi$ gauges, whilst
Eq.(\ref{GGLzero}) is valid at 1-loop but is spoiled by higher order 
corrections. From Eq.(\ref{GGLzero}) and Eq.(\ref{GZLzero}) and from the 
analyticity properties of the two-point functions, it follows that, to all 
orders,
\bea
\label{GGTzero}
A_{\gamma\gamma}(0) &=& 0\\
\label{GZTzero}
A_{\gamma Z}(0) &=& 0~.
\eea
In the $R_\xi$ gauges Eq.(\ref{GGTzero}) is valid at 1-loop, while
Eq.(\ref{GZTzero}) does not hold.
An important  consequence of Eqs.(\ref{QEDWard}-\ref{GZTzero})
is that  in the SM, when the BFM is employed, 
the renormalization  of the electric charge receives contributions only 
from the photon vacuum polarization, analogously to QED. It follows that
the relation between bare charge  and the renormalized one can be written
as in Eq.(\ref{eq1}) and the Dyson summation is justified not only
for the QED part but for the complete SM contribution. Therefore in the
SM the relation between $e_0$ and $e$ is obtained from Eq.(\ref{eq1})
with $\piggp{f}(0)$ replaced by the complete (bosonic plus fermionic) 
$\pigg(0)$ evaluated with the BFM Feynman rules for the SM.
We would like to stress that, differently from the conventional analysis in 
the standard $R_\xi$ gauge, the BFM approach makes manifest the possibility 
of the Dyson summation also for the bosonic part of the vacuum polarization 
function, a fact already discussed in Refs\cite{ddw,dd}.

\section{Results}
\label{results}
Before presenting the result for the two-loop contribution to the vacuum
polarization function we briefly discuss some interesting aspects of a 
two-loop BFM calculation. 

The presence of two different kinds 
of fields, the background and the quantum ones, requires the introduction of
two different sets of Feynman rules, one for the quantum fields that are
actually identical to conventional ones, and one for vertices where 
at least one background field is present. Since the gauge-fixing term
of Eq.(\ref{BFG})  differs from the 
conventional one, Eq.(\ref{RFG}),  by terms that involve both classical
and quantum fields, the corresponding mixed vertices are modified. 
In particular, because  Eq.(\ref{BFG}) is quadratic in the quantum fields,
only vertices in which two quantum fields are present can differ from the
conventional ones, like for example the ${\hat \gamma}W^+ W^-$ vertex
that acquires a $\xi_Q$ dependence. Furthermore, the non linearity of the 
gauge-fixing function  induces a modified ghost sector with respect to 
the linear $R_\xi$ gauges. In a two-loop calculation both sets
of Feynman rules are needed. In fact, in the case of the electric charge,
the external photon is a  background field and couples to the 
bosonic particles running into the loop  differently from an  internal photon,
which instead 
should be regarded as a  quantum field. A complete set of BFM Feynman 
rules can be found in Ref. \cite{ddw}.

The QED-like BFM identities  simplify considerably the renormalization 
procedure. Indeed, it is convenient to choose  a renormalization prescription
that automatically respects  Eqs.(\ref{QEDWard}-\ref{GZTzero}) and for our
two-loop calculation this should be enforced at the one-loop level.
Possible subtleties of this implementation are only related to the
bosonic sector. 
We recall that in the one-loop diagrams, besides the fermions for which we 
employ the usual on-shell mass renormalization,  the particles that
contribute to the bosonic part of the vacuum polarization, $\piggp{b}(0)$,
are the $W$ boson, its unphysical counterpart, and the charged ghosts, 
whose masses squared are $\mw$ and $\xi_Q \mw$ respectively.  
It is then clear that if we renormalize
the masses of all these particles in the same way, namely employing the
same $W$ mass counterterm, $\delta \mw$, for all,   
Eqs.(\ref{QEDWard}-\ref{GZTzero}), that are satisfied at one-loop,
will be automatically preserved under renormalization. This choice
corresponds to employing a gauge fixing function written in terms of
bare parameters and fields.
The tadpole contribution needs a detailed comment. We perform the
standard tadpole subtraction, namely we choose the tadpole counterterm
to cancel the complete one-loop tadpole contribution. This induces an
additional term in the mass counterterm of the unphysical scalar proportional
to one-loop tadpoles. This contribution is needed to restore a topology of 
two-loop diagrams canceled by our choice of the tadpole counterterm and does
not invalidate the preservation of the QED-like Ward identity under our 
renormalization prescription. 

Several other prescriptions for the renormalization of the gauge fixing part 
and associated ghost sector are conceivable. In particular, one can add the 
gauge-fixing term to the renormalized Lagrangian, so that Eq.(\ref{BFG}) is 
expressed in terms of renormalized quantities. In this case, while the mass
of unphysical scalar is not renormalized, a part from the tadpole contribution,
the counterterm of the charged ghost mass becomes 1/2 that of $W$ boson.
However,  besides a counterterm for the $W$--$\phi$ transition, 
several new contributions involving coupling and mass counterterms are 
induced due to the mismatch between the bare  quantities appearing in the
classical lagrangian and the renormalized quantities in the gauge-fixing
term. We have explicitly verified that the two procedures give the same
result. Furthermore we have also explicitly verified the two identities,
Eq.(\ref{QEDWard}) and Eq.(\ref{GZTzero}), at the two-loop
level  in the BFM Feynman gauge, $\xi_Q=1$. 

The BFM allows to write the relation between the bare and renormalized
electric charge as:
\bea
    e^2 &=& {e^2_0 \over 1 - e^2_0\,\pigg(0) }~,
\label{eq1SM} \\
\pigg(0) &=& \piggp{f}(0) + \piggp{b}(0)~,
\label{pis}\\
\piggp{f}(0) &=& \piggp{l}(0)  +  \piggp{p}(0) + \piggp{5}(0) \nonumber\\
             &=& \piggp{l}(0)  +  \piggp{p}(0) + 
              \left(\piggp{5}(0) - {\rm Re}\, \piggp{5}(\mzs)\right) +
               {\rm Re}\,\piggp{5}(\mzs)~,
\label{pif}
\eea
where the fermionic contribution has been separated into a leptonic part,
$\piggp{l}$, a perturbative quark contribution, $\piggp{p}$, and a 
non-perturbative one, $\piggp{5}(0)$. The latter, associated to
diagrams in which a light quark couples to a photon, can be 
related to  $\Dah \equiv  4 \pi \alpha 
\left({\rm Re}\,\piggp{5}(\mzs) -\piggp{5}(0)\right)$ 
that can be evaluated from the experimental data on the cross 
section $e^+ e^- \rightarrow hadrons$ by using a dispersion relation\footnote{
For an alternative approach that evaluates directly $\piggp{5}(0)$ via an 
unsubtracted dispersion relation see Ref.\cite{erl}.} while the other term, 
${\rm Re}\,\piggp{5}(\mzs)$, can be analysed perturbatively.
The  top contribution to the vacuum
polarization can be reliably calculated in perturbation theory because
of the large value of the top mass. Similarly, 
two-loop   diagrams in which a light quark  couples internally
to the $W$ and $Z$ bosons allow a perturbative
evaluation. These contributions  together with the
top ones are  collected in  $\piggp{p}(0)$. 

We report here the one and two--loop irreducible 
perturbative contribution to the BFM photon vacuum polarization 
function evaluated at zero momentum transfer, with the one-loop
result expressed in terms of the physical masses of the fermions and of the
$W$ boson.  We express all the results in units
$1/(16 \,\pi^2)$.
The leptonic part is given by
\bea
\piggp{l}(0) &=& \frac{I_l}{(n-4)}+ \sum_l \left\{\frac43
    \log ({m_l^2\over \mu^2}) \left(1 + \frac{3\alpha}{4\pi}\right) -
     \frac{15\, \alpha}{4 \pi} \right. \nonumber\\
&-& \left.  \frac{\alpha}{4 \pi s^2} \left[ {151 \over 36 } -
       {13 \over 3} \log ({\mw\over \mu^2}) 
+   \frac{1}{c^2} \left(\frac14 - s^2 + 2 s^4\right)  
    \left(\frac32 - 2  \log ({\mzs\over \mu^2} \right) \right] \right\} 
\label{finpil}
\eea
where $m_l$ are the lepton masses.\\
The perturbative quark contributions, including QCD corrections, is
given by ($\zt \equiv \mzs/\mt,\: \htt \equiv \mh/\mt,\:\tw \equiv \mt/\mw$)
\bea
\piggp{p}(0) &=&  \frac{I_p}{(n-4)} +
   \frac{16}9 \log ({\mt \over \mu^2})\left(1 +\frac{\,\alpha_s}{\pi}+
            \frac{\alpha}{3\,\pi}\right) - \frac{20\,\alpha_s}{3\,\pi}
    - \frac{20\, \alpha}{9 \,\pi} \nonumber\\
&-& N_c\frac{\alpha}{4 \pi s^2}  \left[
\frac{4\,\left(- 17 + 40\,c^2 - 32\,c^4 \right) \,{\zt}^2}{243\,
     c^2\,\left( -4 + \zt \right) }
\right. \nonumber \\
&&+\frac{108 + 
\left( -443 - 800\,c^2 + 640\,c^4 \right) \,\zt + 
     \left( 573 - 840\,c^2 + 672\,c^4 \right) \,{\zt}^2 }{486\,
     c^2\,\left( -4 + \zt \right) \,\zt} \nonumber \\
&&  + 
  \frac{4\,\left( 7 + 17\,\zt - 
       40\,c^2\,\left( 2 + \zt \right)  + 
       32\,c^4\,\left( 2 + \zt \right)  \right) \,
     \, B0(m_t^2,m_t^2,\mzs)}{243\,c^2} \nonumber \\
&&+ 
  \frac{2\,\left( 27 + \left( -37 - 40\,c^2 + 32\,c^4 \right) \,
        \zt + \left( 34 - 80\,c^2 + 64\,c^4 \right) \,
        {\zt}^2 \right) \,\log ({\mt \over \mu^2})}{243\,c^2\,\zt} 
\nonumber \\
&&+ \frac{2\, \log (\zt)}{243\,c^2
{\left( -4 + \zt \right) }^2}
 \left( 126 + 637\,\zt - 275\,{\zt}^2 + 
       34\,{\zt}^3 \right.  \nonumber \\
&& ~~~~~~~~~~~~~~~~~~~~~~~~- 40\,c^2\,\left( 36 + 20\,\zt - 
          13\,{\zt}^2 + 2\,{\zt}^3 \right)  
\nonumber \\
&& ~~~~~~~~~~~~~~~~~~~~~~~~\left. 
       +32\,c^4\,\left( 36 + 20\,\zt - 13\,{\zt}^2 + 
          2\,{\zt}^3 \right)  \right)  \nonumber \\
&&- 
  \frac{4\,\left( -7 - 40\,c^2\,\left( -2 + \zt \right)  + 
       32\,c^4\,\left( -2 + \zt \right)  + 8\,\zt \
\right) \,\phi(\frac{\zt}{4})}{27\,c^2\,
     {\left( -4 + \zt \right) }^2\,\zt}
\nonumber \\
&&+ \frac{4\,\left( -4 + \htt \right) \,
     \Mvariable{}\,\, B0(\mh,\mt,\mt)}{27\,
     {\Mvariable{c}}^2\,\zt}
+  \frac{2\,\left( -6 - 11\,\htt + 2\,{\htt}^2 \right)
\,
     \Mvariable{}\,\log (\htt)}{27\,
     \left( -4 + \htt \right) \,\zt \, c^2}
\nonumber \\
&& +\frac{ (25 - 8 \,\htt)}{54\, \zt\, c^2}- 
\frac{ (10 -4 \,\htt)}{27\, \zt\, c^2} \log ({\mt \over \mu^2})
      + \frac{4\,\left( -1 + \htt \right) \,
        \phi(\frac{\htt}{4}) } {9\,
     \left( -4 + \htt \right) \,\htt\,\zt\, c^2} \nonumber \\
&&-  \frac{29}{36} 
-\frac{8}{27\, \tw}+ \frac{379}{216} \,\tw +\frac76 \,\tw^2  \nonumber \\
&&- \left(\tw -1 \right) \left(2 + \tw \right) \left(\frac76  
\, B0(\mw,0,\mt) +
\frac4{27 \, \tw}\, B0(\mt,0,\mw) \right) \nonumber \\
&&+ \frac{\tw \left(26+ 7\, \tw - 63\, \tw^2 \right)}{54 (\tw-1)} \log (\tw)
+ \frac{16  -92\, \tw -56\, \tw^2 - 63 \,\tw^3}{54\, \tw}
 \log ({\mw\over \mu^2}) \nonumber \\
&&+ \left. 
\frac1{c^2} \left( \frac{11}{72}-\frac{19}{54} s^2 + \frac{35}{81} s^4 
\right)
\left(3 -  4\,\log ({\mzs\over \mu^2}) \right) + \frac{139}{18} - \frac{70}9
 \log ({\mw\over \mu^2}) \right] 
\eea
\label{finpit}
where in the last line the perturbative contributions 
of the first 5 light quarks is collected.\footnote{The bottom contribution
includes only diagrams with the $Z$ exchange.}

The light quark contribution Re$\,\piggp{5}(\mzs)$
has been discussed in detail in \cite{djve,dg}.
For completeness we report the result:
\bea
{\rm Re}\,\piggp{5}(\mzs) &=& \frac{I_5}{(n-4)} \nonumber \\
&+&
4 \sum_{q\neq t} Q_q^2
\left[ \log ({\mzs \over\mu^2}) \left(1 +
        \frac{\alpha_s}{\pi} + \frac{3 \alpha}{4 \pi} Q^2_q \right)
-\frac53 +\left(\frac{\alpha_s}{\pi} + \frac{3 \alpha}{4\pi} Q_q^2
\right) \left(4 \zeta(3) -\frac{55}{12} \right) \right]~.
\nonumber \\ \  
\eea

Finally, the terms of purely bosonic origin are ($\hw \equiv \mh/\mw$):
\bea
\piggp{b}(0) &=&    \frac{I_b}{(n-4)} -7 \log (\frac{\mw}{\mu^2})+\frac{2}{3} 
\nonumber \\
&-&\frac{\alpha}{4\pi\, s^2}\Bigg\{
 {-7\,\hw^2\,c^4 + 28\, \hw\, c^4 + 109 - 668 \,s^2 + 888 \, s^4
   - 336 \, s^6 \over 12\,{\Mvariable{c}}^4}  \log (\frac{\mw}{\mu^2})
 \,+ \nonumber\\
&&  \frac{\left( -7\,\hw^4 + 77\, \hw^3 - 322\,\hw^2 + 468\,\hw + 
       72 \right) }{12\,
     {\left( \hw - 4 \right) }^2} \,\log (\hw)+ \nonumber\\
&&  \frac{\left(108 - 1047\,{\Mvariable{s}}^2 + 
       2086\,{\Mvariable{s}}^4 - 1356\,{\Mvariable{s}}^6 + 
       216\,{\Mvariable{s}}^8 \right)}
{12\,{\Mvariable{c}}^4\,
     \left( 1 - 4\,{\Mvariable{c}}^2 \right) } \,\log (c^2) - \nonumber\\
&&  \frac{7\,\left( \hw^2 - 4\,\hw + 12 \right) \,
     }{12}\, {B0}(\mw,\mh,\mw) 
+\frac{3 \left( 3\,\hw - 12 + {4\over \hw} \right)}
   {2\,{\left( \hw - 4 \right) }^2} \,\phi(\frac{\hw}{4}) -\nonumber\\
&&  \frac{7\,\left( -99 + 264\,{\Mvariable{s}}^2 - 
       212\,{\Mvariable{s}}^4 + 48\,{\Mvariable{s}}^6 \right)}{12\,
     {\Mvariable{c}}^4} \,\, {B0}(\mw,\mw,\mzs)
- \frac{9\,{\Mvariable{c}}^2\,
     \left( 3 - 4\,{\Mvariable{s}}^2 + 4\,{\Mvariable{s}}^4 \right)} 
     {2\left( 1 - 4\,{\Mvariable{c}}^2 \right) }\,
     \phi(\frac{1}{4 \Mvariable{c}^2})+\nonumber \\
&& \frac{1}
{36\,c^4\, \left( \hw - 4 \right) } 
\Big[21\,{\Mvariable{c}}^4\,\hw^3 -
  153\,{\Mvariable{c}}^4\,\hw^2 + \nonumber\\
&&      \hw\,\left( -379 + 3464\,{\Mvariable{s}}^2 - 
        5404\,{\Mvariable{s}}^4 +
        2340\,{\Mvariable{s}}^6 \right)  +
4\,\left( 664 - 4034\,{\Mvariable{s}}^2 +
        5689\,{\Mvariable{s}}^4 - 
        2340\,{\Mvariable{s}}^6 \right) \Big]
\Bigg\} \nonumber \\
\ 
\label{finpib}
\eea
The divergent parts of 
$\piggp{i}$   denoted by $I_i ~(i=l,p,5,b)$
are, in units $1/(16 \pi^2)$:
\bea
I_l&=& 
\sum_l \left[
\frac83 +\frac{\alpha}{4 \pi s^2} 
 \left(   
4 \, s^2 + \frac{13}{3} + \frac{1}{2\,c^2} (1-4 s^2 + 8 s^4)
\right) 
\right]
\\
I_p&=& N_c 
\left[
\frac{32}{27} \left(1 +\frac{\alpha_s}{2\,\pi}+
            \frac{\alpha}{6\,\pi} \right) +
\frac{\alpha}{4 \pi s^2}
\left(-\frac{13}{18}\frac{\mt}{\mw} +\frac{255 -318 s^2 +136 s^4}{54\, c^2}
\right)
\right]
\\
I_5&=& \frac{44}9 \left(2 +  \frac{\alpha_s}{\pi} \right) +
        \frac{35\,\alpha}{27\, \pi}
\\
I_b&=& 
- 14 - \frac{\alpha}{4 \pi s^2} \frac{125-128 s^2}{6 c^2} 
\label{divpib}  
\eea
In Eqs.(\ref{finpil}-\ref{finpib}) $B0(s,m_1,m_2)$ is the real part of the 
scalar 1-loop self-energy integral defined as:
\be
\label{Bzero}
B0(s,m_1,m_2) = -\int dx \log \frac{x^2 s -x(s+m_1-m_2) + m_1}{\mu^2}
\ee
whose explicit expression can be found, e.g, in \cite{degsir} and
\be
\phi(z) = \left\{
       \begin{array}{ll}
       4 \sqrt{{z \over 1-z}} ~Cl_2 ( 2 \arcsin \sqrt z ) &  0 < z \leq 1\\
       { 1 \over \lambda} \left[ - 4 {\rm Li_2} ({1-\lambda \over 2}) +
       2 \log^2 ({1-\lambda \over 2}) - \log^2 (4z) +\pi^2/3 \right]
       & z >1 \,,
       \end{array} \right.
       \label{e2.15c}
\ee 
where $Cl_2(x)= {\rm Im} \,{\rm Li_2} (e^{ix})$ is the Clausen function
and $\lambda = \sqrt{1 -1/z}$.

The on-shell two-loop electric charge counterterm, $ 2 \delta e^{(2)}/e$,
is given by the two-loop contribution to the BFM photon vacuum polarization
function, namely the terms explicitly proportional to $\alpha$ (or
$\alpha_s$) in Eqs.(\ref{finpil}-\ref{divpib}). We stress that 
 $ 2 \delta e^{(2)}/e$ is a gauge invariant quantity that does not depend
on the gauge fixing procedure employed to compute it. 

To check our results we have computed the two-loop amplitude to the
Thomson scattering in two different ways. First, employing the BFM
gauge-fixing procedure assuming $\xi_Q=1$. In this case the amplitude is 
directly proportional to $J_\gamma$ through:
\be
\label{twoloopBFG}
{\cal M}^{(2)}_{BFM} = 
\frac{1}{2\, q^2} \agg^{(2)}(0) ~+~
\frac{3}{8 \,q^4} \agg^{(1)}(0) \agg^{(1)}(0)
\ee
where the factors $3/8$ and $1/2$ take into account the wave function
renormalization of the external photon and the superscript (1,2) indicates
the loop order. 

In the second case we have used the conventional $R_\xi$ gauge-fixing 
procedure with $\xi=1$. In this case the  vertex
corrections are different from zero\footnote{We include in the vertex
corrections also the wave function renormalization of the external fermions.}
and give rise to two contributions, proportional to $J_{\gamma}$ and
to $J_Z$ respectively. Accordingly, the total amplitude is composed
by two parts, one proportional to the photonic current, 
${\cal M}^{(2)}_{R_\xi,J_{\gamma}}$,  while the other proportional to
the $Z$ current, ${\cal M}^{(2)}_{R_\xi,J_{Z}}$. Calling 
$V_{\gamma,~J_{\gamma}}^{(i)} \:(V_{\gamma,~J_Z}^{(i)})$, ($i=1,2$) the part
of the photon vertex proportional to $J_{\gamma}$ ($J_Z$) and analogously
for the $Z$ vertex we have:
\bea
\label{twoloop1}
{\cal M}^{(2)}_{R_\xi, J_\gamma} &=& 
\frac{1}{2 \,q^2} \agg^{(2)}(0) + 
\frac{-1}{2 \,q^2 \,\mzs} \agz^{(1)}(0) \azg^{(1)}(0) +
\frac{3}{8\, q^4} \agg^{(1)}(0) ~\agg^{(1)}(0) +
\nonumber \\
&&V_{\gamma,~J_{\gamma}}^{(2)} +
\frac{1}{2\, q^2}  V_{\gamma,~J_{\gamma}}^{(1)}~\agg^{(1)}(0) +
\frac{-1}{\mzs} V_{Z,~J_{\gamma}}^{(1)}~\azg^{(1)}(0)~,
\eea
\bea
\label{twoloop2}
{\cal M}^{(2)}_{R_\xi, J_Z} &=& 
\frac{-1}{\mzs} \azg^{(2)}(0) +  
\frac{-1}{2\,\mzs\,q^2 } \azg^{(1)}(0)~\agg^{(1)}(0) + 
\frac{1}{(\mzs)^2} \azz^{(1)}(0)~\azg^{(1)}(0)  + 
\nonumber \\
&&V_{\gamma,~J_Z}^{(2)} +
\frac{1}{2\, q^2} V_{\gamma,~J_Z}^{(1)}~\agg^{(1)}(0) +
\frac{-1}{\mzs} V_{Z,~J_Z}^{(1)}~\azg^{(1)}(0)~.
\eea
We have verified that ${\cal M}^{(2)}_{BFM} ={\cal M}^{(2)}_{R_\xi, J_\gamma}$.
To achieve this the two-loop vertex corrections 
$V_{\gamma,~J_{\gamma}}^{(2)}$ are needed.
To shortcut the calculation one notices that 
$V_{\gamma,~J_{\gamma}}^{(2)}=1/s^2\, V_{\gamma,~J_Z}^{(2)}$ because the
photon vertex should be proportional to $J_3$. The part of the photon
vertex proportional to $J_Z$ can be obtained from Eq.(\ref{twoloop2}) 
since the conservation of the electric charge requires 
${\cal M}^{(2)}_{R_\xi, J_Z} =0$. We recall that
at the 2-loop level, in the 't Hooft Feynman gauge, Eqs.(\ref{GGTzero})
and (\ref{GZTzero}) are not valid. In fact the two terms in
\be
\nonumber
\frac{1}{2 \,q^2} \agg^{(2)}(0) +  
\frac{-1}{2\,\mzs \,q^2} \agz^{(1)}(0) \azg^{(1)}(0)
\ee
show individually a $1/q^2$ pole when $q^2 \rightarrow 0$.  However,
they  cancel each 
other so  that the total amplitude is regular at $q^2=0$.
\section{The Parameter $\ehs(\mz)$}
\label{secehs}
The relation given by Eq.(\ref{eq1SM}) allows to determine one
of the fundamental parameter of the \ms\ renormalization scheme,
$\ehs(\mz)$, i.e. the \ms\ electric charge defined  at scale $\mz$. 
The  \ms\ renormalization procedure is  defined  as the 
subtraction of pole terms of the form $(n-4)^{-m}$, where $m$ is an integer 
$\geq 1$, and the identification
of the 't~Hooft parameter $\mu$ (actually the rescaled one of Eq.(\ref{E5b}))
with the relevant mass scale, in this case $\mz$. One can slightly modify
this basic procedure by implementing the decoupling of heavy particles
\cite{mr,fks}, namely by absorbing  the contribution of particles with mass
greater than  $\mz$ in the definition of  $\ehs(\mz)$, in particular
the contribution of $m_t$. At the two-loop level $\ehs(\mz)$ contains
also a dependence on $m_h$, whose 95\% C.L. direct search lower limit, 
$m_h > 114.4$ GeV, is greater than $\mz$. However, because both the top and 
the Higgs are partners of isodoublets, their ${\cal O}(\alpha^2)$ decoupling
requires a specific matching procedure between the two theories 
above and below their mass values. In the present paper we do not implement
the decoupling of heavy particles.

\begin{table}[t]
\centering
\begin{tabular}{|c|c|c|c|c|}
\hline
              &1loop    &2loop QCD  &2loop QED &2loop EW full   \\
\hline
leptons      & 3529.2   &       &  7.66    & 10.18   \\
\hline
bosons       &   -140.7   &       &          &  -1.79   \\
\hline
top          &   -133.7   & 8.66  &  0.19   &  0.08       \\
\hline
$\piggp{5}(0)\Big|_{EW}$ &       &      &         &   4.56   \\
\hline
${\rm Re}\, \piggp{5}(\mzs)$ & 473.4     & -2.39      &  -0.04  &        \\
\hline\hline
$\Dah$     &  2757.2   &  \multicolumn{3}{c|}{ }        \\
\hline\hline
total        & 6485.4   &  6.27     &    7.81      &     13.03    \\
\hline
\end{tabular}
\caption{Numerical results for $\Delta \alphah (\mzs)$, expressed
  in units $10^{-5}$.
The input parameters are specified in the text.
Different perturbative contributions are presented. }
\label{numbersmz}
\end{table}

In order to obtain the relation between \ehs\ and $e^2$, one
writes $e_0^2 = \ehs/\hat{Z}_{e}$ in Eq.~(\ref{eq1SM}), and uses the
counterterms present in $\hat{Z}_{e}$ to cancel the $(n-4)^{-1}$ terms
in the regularized but unrenormalized vacuum polarization function
$\pigg(0)$  setting $\mu = m_Z$ in the explicit expressions (see 
Eqs.(\ref{finpil}--\ref{finpib})). 
Without implementing any decoupling we have
\be
  \hat{Z}_e = 1
 +{\alphah\over4\,\pi}\left(I_l+I_t +I_5 + I_b\right){1\over n-4}
 \label{E7d}
\ee
so that
\be
  e^2 = {\ehs (\mz) \over1 + (\alphah /\alpha)\Delta \alphah (\mzs)} ,
\label{E8a}
\ee
with 
\bea
\label{E8b}
  \Delta \alphah (\mzs)&=& -4\, \pi \alpha
\left[ \piggpms{l}(0) + \piggpms{p}(0) + \piggpms{b}(0) \right] \nonumber \\
    & & + \frac{\alpha}\pi \left[ \frac{55}{27}
        + \left(\frac{11\alphah_s(\mzs)}{9\pi}
 + \frac{35\alphah(\mzs)}{108\pi}\right) 
   \left(\frac{55}{12} - 4 \zeta(3) \right) \right]
+ \Dah
\eea
where  $\piggpms{i}$ is
the self-energy expression subtracted of its divergent $\frac{I_i}{n-4}$
term with $\mu$  set equal to $ \mz$.

Eq.(\ref{E8b}) can be easily solved  for \ehs, obtaining
\be
  \ehs (\mz)= {e^2\over1 - \Delta \alphah (\mzs)} .
\label{E8c}
\ee

The determination of $\ehs(\mz)$ requires the specification of the
hadronic contribution $\Dah$. Several evaluations of this important
parameter have been presented over the last fifteen years \cite{allworld}.
In our numerical analysis we use the recent determination by
Jegerlehener\cite{jegerlehner}
\be 
\Dah = 0.027572\pm0.000359
\label{fred}
\ee
that together with the following values (in GeV) for the fermion masses
$
m_e = 0.000511, m_{\mu} = 0.105658  , m_{\tau} = 1.777 , m_{t} = 174.3
$
and for the gauge bosons $\mz = 91.187,  
\:\mw = 80.43$ yield, for $m_h = 150$,
$\Delta \alphah (\mzs)= 0.06505\pm 0.00036 $ corresponding to 
$\alphah^{-1} = 128.122 \pm 0.054 $.

In table \ref{numbersmz}
we present separately the various contributions to $\Delta \alphah (\mzs)$.
The perturbative contribution of the first 5 light quarks has been
indicated by $\piggp{5}(0)\Big|_{EW}$.
The different contributions are shown at the 1- and at the
2-loop level.
In the latter case, the QED and QCD contributions were already 
discussed in \cite{fks}.
We have checked, in the lepton and in the top case, that the
appropriate subset of diagrams from our result 
reproduces the numbers presented in \cite{fks}.
Concerning the 2-loop EW diagrams involving a top quark, approximate
results including all terms of order ${\cal O}(\alpha^2 \mt/\mw)$
were already available \cite{DG} and could also be reproduced.\\
The largest contributions are due to light fermions (leptons and
quarks) exchanging massive vector bosons and have both positive sign.
In contrast the 2-loop purely bosonic diagrams have  negative sign
and are smaller in size. Their contribution grows, in absolute value,
with $m_h$ but remains always small: for $m_h = 400$ GeV it reaches
-2.57 in units $10^{-5}$. The top quark contributions deserve a
detailed comment. The inclusion of the full 2-loop EW corrections
makes the result tiny, canceling to a large extent the 
${\cal O}(\alpha^2 \mt/\mw)$ part.
In fact, the expansion of the 2-loop EW corrections in powers of $\mt$
is sensible asymptotically \cite{dgv}, for very large values of
$m_{\scriptstyle t}$; only in this regime, when the top Yukawa coupling is
much larger than the gauge couplings, the terms ${\cal O}(\alpha^2
\mt/\mw)$ are a good approximation of the full results.
In contrast, for realistic values of $m_{\scriptstyle t}$, the
``subleading'' terms are as large as the leading ones and can not be
neglected. The fact that a large cancellation occurs should be
considered fortuitous.\\
The size of the full 2-loop EW results is more than 10 parts in units
$10^{-5}$ and almost half of it is due to purely electroweak effects.
These results are  comparable to the error given in the so called 
``theory-driven'' analyses of $\Dah$ which yield, for instance
$\Dah = 0.02763\pm 0.00016$ \cite{dh}.

\begin{table}[t]
\centering
\begin{tabular}{|l|c|c|c|c|c|}
\hline
$\mu$ [GeV] & 1loop +NP   & 2loop QCD & 2loop EW full & total & 
   $\hat\alpha^{-1}(\mu)$  \\
\hline
~~91.187 & 6485.42 &  6.27 & 13.03 & 6504.72 & 128.122 $\pm$ 0.054 \\
\hline
~300     & 6991.91 & 40.90 & 21.45 & 7054.26 & 127.369 $\pm$ 0.054 \\
\hline
~500     & 7209.15 & 55.75 & 25.05 & 7289.96 & 127.046 $\pm$ 0.054 \\
\hline
~800     & 7409.01 & 69.42 & 28.37 & 7506.81 & 126.748  $\pm$ 0.054 \\
\hline
1000     & 7503.90 & 75.91 & 29.94 & 7609.76 & 126.607  $\pm$ 0.054 \\
\hline
\end{tabular}
\caption{
Numerical results, in  units $10^{-5}$ for 
$\Delta \alphah (\mzs)$ for different values of $\mu$.
In the first column the non-perturbative hadronic contributions  
is added to the 1-loop results. }
\label{numbershe}
\end{table}
The gauge invariant inclusion of the bosonic contributions in the
definition of the effective running coupling is relevant when we
consider high-energy processes, like the ones that will be studied at
the LHC or at TESLA.
In the table \ref{numbershe} we present the value of $\ehs(\mu)$ for 
$\mu=300, \,500, \,800, \,1000$ GeV. We employ the same value for the
hadronic contributions, i.e. Eq.(\ref{fred}),
and include the full one- and two-loop results for the perturbative
part.

\section{Conclusions}
\label{concl}
We  presented the results of the calculation of the complete
2-loop electroweak corrections to the Thomson scattering amplitude,
which allowed us to fix the electric charge counterterm in the
on-shell scheme.

We emphasized the advantages offered by the BFM for the quantization
of the theory, both from the theoretical and from the computational
point of view. In particular, the BFM makes manifest the possibility
of Dyson summation for the complete photon vacuum polarization
function. 

We studied the effective $\ms$ coupling $\ehs(\mu)$
and evaluated it numerically for different values of the scale $\mu$.
In particular, for  $\ehs(\mz)$, the effect of the 2-loop EW corrections is 
twofold:
$i)$ they shift the central value and 
$ii)$ reduce the theoretical perturbative uncertainty on its determination, 
which is now pushed at the 3-loop level.
Concerning the first point, the indirect Higgs boson mass determination 
from a global fit to all electroweak precision observables is very 
sensitive to the precise input value
for $\ehs(\mz)$. In fact, a variation of the central value of $\ehs(\mz)$
by $5 \cdot 10^{-5}$, that can be taken as the difference between
the value of  $\ehs(\mz)$ determined including the complete
two-loop electroweak corrections and that obtained including only
the two-loop QED part,  gives a reduction in the 95 \% upper limit for
the Higgs mass ${\cal O}(6$-$8)$ GeV.
\section*{Acknowledgments}
This work was partially supported by the European Community's
Human Potential Programme under contract
HPRN-CT-2000-00149 (Physics at Colliders).

\end{document}